\documentclass[twocolumn,showpacs,superscriptaddress,10pt]{revtex4-1} 
\usepackage{natbib,amsmath,amssymb,graphicx}

\begin{document}
\draft
\title {Nonlinear coupler operating on Werner-like states -- entanglement creation, its enhancement and preservation}
\author {A. Kowalewska-Kud{\l}aszyk}
\affiliation{Nonlinear Optics Division, Faculty of Physics, Adam Mickiewicz
 University, Umultowska 85, 61-614 Pozna\'n, Poland}
\author {W.Leo\'nski}\email{Corresponding author: w.leonski@if.uz.zgora.pl}
\affiliation {Quantum Optics and Engineering Division, Institute of Physics, University of Zielona G\'ora, Z. Szafrana 4a, 65-516 Zielona G\'ora, Poland}


\begin{abstract}
We discuss a model of two nonlinear Kerr-like oscillators, mutually coupled and excited by parametric process. We show that the system's evolution, starting from Werner-like states, remains closed within a small set of two-mode n-photon states the system, and pure two-qubit entangled state can be generated. For some initial Werner-like states delayed entanglement generation can be observed. We investigate the influence of two damping mechanisms on the system's evolution. We show that for the both cases, the entanglement can survive despite the presence of damping, and the effects of sudden entanglement death and its rebirth can appear in the system.
\end{abstract}


\maketitle
\section{Introduction}
Evolution of quantum entanglement and its decay seem to be one of the most striking problems, frequently discussed in recent years. Especially, subject of sudden disappearance of the entanglement (\textit{sudden entanglement death}) and its revival is located inside very interesting groups of problems related to sudden vanishing and reappearance of nonclassicality or quantum correlations.

In this paper we shall concentrate on the evolution of Werner (or Werner-like states) \cite{W89,MJW01,GKS01,WNG03} in a system of two mutually coupled quantum Kerr-like oscillators. Such nonlinear oscillatory system describes for instance, fiber coupler systems used as optical switching devices \cite{ZZ12a,ZZ12b,Z13}. What should be emphasized, Kerr-like models that are discussed here could also be implemented  in other, not necessarily optical systems. For instance, there could be  Bose-Einstein condensate models \cite{PLK13} or  nanomechanical resonators \cite{YMG10}. The systems involving Kerr-like quantum oscillators can also be applied as sources of quantum states defined in finite-dimensional Hilbert space, and are referred to as \textit{nonlinear quantum scissors} (NQS). Applying such systems one can generate one-mode  \cite{LT94,AMPS00,GSK12,GK13b,MPL13}, two-mode \cite{LM04,KL06}, or three-mode truncated states \cite{SWU06} (for review of problems related to NQS see \cite{MLI02,LM02,LK11} \textit{and the references quoted therein}. From other side, one should mention that  Werner or Werner-like states that we shall discuss here, are widely applied in various quantum information theory models. For instance, they were applied in systems in which entanglement teleportation was realized \cite{LK00}, or used as entanglement witnesses \cite{BMN03}.  Moreover, Werner states were discussed in a context of violation of Bell inequality \cite{M04}.

In particular, we shall discuss here time-evolution of Kerr-like coupler which was initially in the Werner-like state.  We will show that our system, interacting with external bath, behaves as two-qutrit one.
Although the evolution of the entanglement in two-qubit, qubit-qutrit or two-qutrit systems, interacting with various types of reservoirs, was already discussed in the previous papers (for instance, see \cite{ML04,KL06,KL09,KLP11} \textit{and the references quoted therein}) but was considered only for cases of initial pure states: Fock or maximally entangled states. One of the purposes of this paper is to show that effective state truncation and entanglement creation is also possible for other initial states: mixed states and Werner-like ones (which are mixtures of Bell and mixed states). Moreover, we will show that if we assume initial Werner state, it is possible to observe not only SED but also delayed birth of the entanglement. 
In that  way, we would like to  emphasize that the effects considered here are  not restricted to the cases of specially chosen initial conditions which were discussed in previous papers ($n$-photon states with carefully chosen number of photons). What is important, Werner-like states contain some amount of noise and hence, are easier to achieve experimentally than pure $n$-photon states.

The paper is organized as follows: at first we discuss in more details our model of NQS  and define its initial conditions. After that, we describe the process of two-mode state truncation and concentrate on the system's time-evolution, discussing the dynamics of entanglement for two-qubit subsystems. We shall show that our model, starting its evolution from mixed Werner-like state, can lead to pure entangled states generation. In particular, delayed entanglement creation from initially not-entangled states can be observed. Then, we concentrate on the influence of two mechanisms of damping processes (phase and amplitude damping) on the entanglement dynamics, showing that for the both of them, depending on the parameters describing initial Werner-like state, asymptotic decay of entanglement or its sudden death (and rebirth) can appear in the system.  Moreover, we will derive and discuss analytical conditions necessary for the appearance of such phenomena.

\section{The model and its initial states}

We concentrate here on the nonlinear systems involving two nonlinear quantum oscillators described by third-order susceptibilities $\chi$. Such nonlinearitites can also be referred to as Kerr-like ones. In fact all systems which we describe by effective Kerr-like Hamiltonians can represent various physical models, for example cavities with vibrating mirrors \cite{MMT97,BJK97}, condensed matter systems \cite{KDZ09} or others. Therefore, our considerations not necessarily concern  systems involving usual optical Kerr effect.

For the model discussed here the both: interaction between two nonlinear parts and the external excitation are supplied by a parametric process of strength $g$. Such system can be realized as it was proposed in \cite{KLP11}. In particular, two photons (of two separate field modes) which are produced in parametric down-convertion process (PDC)  interact with two nonlinear systems described by the Kerr-like nonlinearities. In that way the entanglement in polarizations (which is the feature of photons pairs produced in PDC process) is converted into the entanglement in photon number states.

The Hamiltonian of the system can be written as follows:
\begin{equation}
\hat{H}=\underbrace{\frac{\chi_a}{2}(\hat{a}^\dagger)^2\hat{a}^2+
\frac{\chi_b}{2}(\hat{b}^\dagger)^2\hat{b}^2}_{\hat{H}_{NL}}+\underbrace{g\hat{a}^\dagger\hat{b}^\dagger +g^{\star}\hat{a}\hat{b}}_{\hat{H}_{par.}}
\label{heq1}
\end{equation}
where $\chi_{a(b)}$ are the Kerr-like nonlinearities, whereas $\hat{a}^{+}(\hat{b}^{+})$ are photon creation operators corresponding to field modes $a$ and $b$, respectively.

In previous considerations where parametrically pumped nonlinear coupler \cite{KLP11} was discussed, initial states were assumed to be pure number states (for example vacuum state in both modes, or one-photon states in both modes). It was shown there  that for such cases treating  the system as NQS is fully justified. For such system formation of two-qubit or two-qutrit entangled states where described and  dynamic flow of entanglement between different two-qubit subspaces where also spotted. 
It was found that in one of the system's subspaces entanglement vanishes in a finite time after some oscillations are observed. In the other subspace, entanglement decays and after a significantly long time revives there. In the last of these two subspaces the entanglement after some oscillations survives at some constant value. 

Here,  we extend the group of initial states for our model and use the mixtures of maximally entangled states and  maximally mixed states. Such mixture of the states is called Werner \cite{W89}, or Werner-like \cite{MJW01,GKS01,WNG03} ones. We shall show how such states influence Hilbert space truncation scheme (effectiveness of quantum scissors) and formation of entangled two-qubit states. In particular, we use states defined in a $d^n$ dimensional space as:
\begin{equation}
W^{[d^n]}(s)=(1-s)\frac{1}{d^n}\mathcal{I}^{(n)}+s|\Psi\rangle\langle\Psi |
\label{wer1}
\end{equation}
where:
$\mathcal{I}^{(n)}$  is the identity matrix in the $d^n$ dimensional Hilbert space; matrix
$\frac{1}{d^n}\mathcal{I}^{(n)}$  corresponds to the  maximally mixed state, whereas
$|\Psi\rangle\langle\Psi |$  is density matrix for maximally entangled pure  state (MEPS) component. The parameter $s\in\langle 0,1\rangle$ describes the probability of contribution of MEPS in initial state. Depending on the value of this probability, Werner states are either separable or entangled \cite{BBP96,BCJ99}. In fact, in our considerations, as we deal with two-qubit subspaces,  we use Werner-like states with $n$ and $d$ equal to $2$.
As MEPS we use Bell states. They are:
\begin{subequations}
\begin{eqnarray}
&|B\rangle_1&=\frac{1}{\sqrt{2}}\left(|0\rangle_a|0\rangle_b \pm |i\rangle_a|i\rangle_b\right)\hspace*{5em}\\
\hspace*{-4cm}\mbox{or}&&\nonumber\\
&|B\rangle_2&=\frac{1}{\sqrt{2}}\left(|0\rangle_a|i\rangle_b \pm |i\rangle_a|0\rangle_b\right)\label{bell-like}
\end{eqnarray}
\end{subequations}
for $i=1,2$.

\section{Space Truncation}
We start our considerations from the case of non-dissipative system. To determine the system's evolution, we construct unitary evolution operator applying the Hamiltonian (1) in a form of matrix defined in $n$-photon states basis. Then, analogously as in \cite{L96}, we act using this evolution operator on the initial state to get the wave-function describing the system for arbitrary moment of time. This procedure can be easily done numerically and its technical details can be found for instance, in \cite{NR2007, DLC13}. 

For initial two-mode Fock states with equal number of photons in each mode, the number of two-mode states involved within the dynamics of the whole system is limited to the three states only. It can be achieved for moderate values of parametric pumping strength, when compare it with nonlinearity constants values. In consequence, our system behaves  as a two-qutrit one. This situation is analogous to that discussed in \cite{KLP11} and, quite recently, in \cite{GSCK13,GK13a}.

The fact that mixed states are added to the MEPS leads to increasing number of two-mode states that influence the system's dynamics. Their probability amplitudes depend on the form of the pure maximally entangled state and on the degree of its contribution in Werner-like initial state. It is clear that the larger the  value of $s$, the more the system resembles two-qutrit one. But anyway, even when mixed state dominates over the entangled component, there is a nonzero probability for generation of entanglement within all of the two-qubit subspaces.

\begin{figure}[htbp]
\centerline{\includegraphics[width=1\columnwidth]{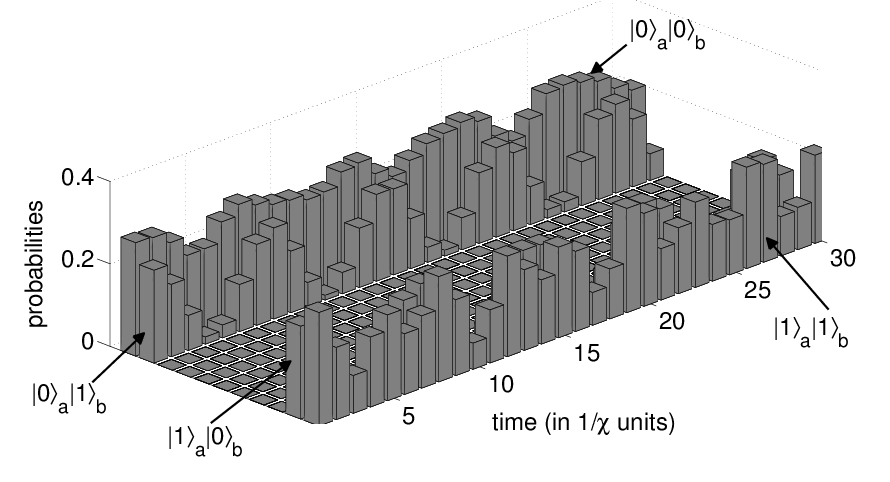}}
\centerline{\includegraphics[width=1\columnwidth]{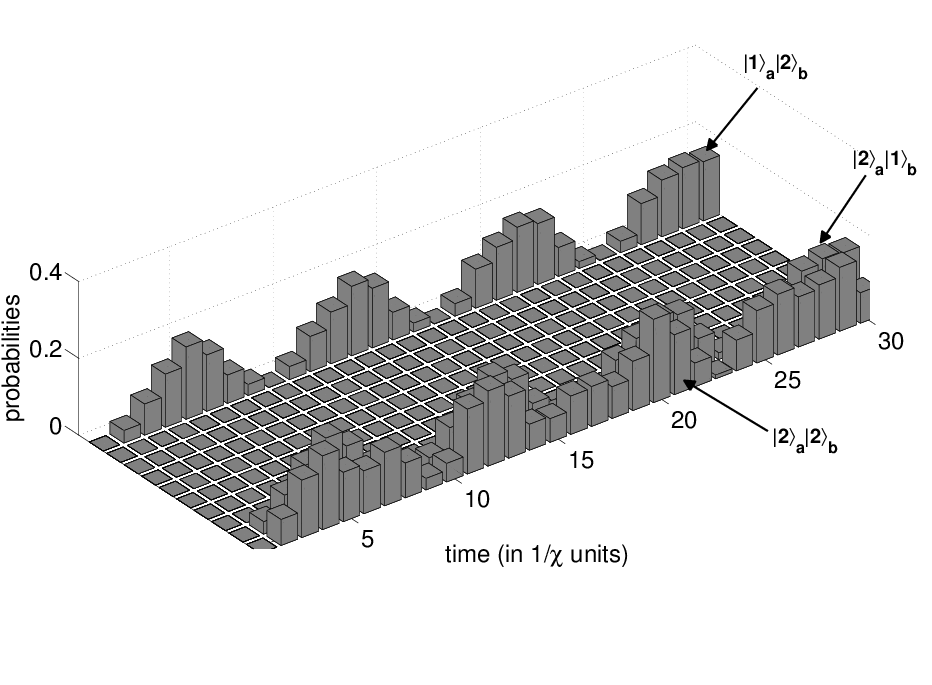}}
\caption{Probability amplitudes for the states involved in the system's evolution and for no-damping case. We assume that initial state is Werner-like state (2) with $s=0.1$ and $|B\rangle_1=\frac{1}{\sqrt{2}}\left(|0\rangle_a|0\rangle_b \pm |1\rangle_a|1\rangle_b\right)$. We assume that  $g=0.6\chi$ and $\chi_=\chi_b=\chi$=1. Time is scaled in units of $1/\chi$.}
\end{figure}

Thus, Fig.1 shows the situation when Bell state $|B\rangle_1$ (with $i=1$) 
is chosen as the MEPS component of initial state. In practice, for small values of $s$ (like $s=0.1$) only four two-mode states ($|0\rangle_a|0\rangle_b$, $|0\rangle_a|1\rangle_b$, $|1\rangle_a|0\rangle_b $ and $|1\rangle_a|1\rangle_b$) are involved in the system's dynamics.  However, as a result of parametric interaction, apart from these states (which have created two-qutrit system for the case of initial  Fock state) also 2 other states: $|1\rangle_a|2\rangle_b$, $|2\rangle_a|1\rangle_b $  which are coupled to initially populated ones $|0\rangle_a|1\rangle_b$, $|1\rangle_a|0\rangle_b $,  are involved in the system's evolution.
Although the initial states belong to one group of states defined in  subspace $\{0110\}$, during the evolution of the system these  states can be divided into two decoupled groups ($|0\rangle_a|0\rangle_b$, $|1\rangle_a|1\rangle_b$ and $|0\rangle_a|1\rangle_b$, $|1\rangle_a|0\rangle_b$). This is a consequence of the fact that the parametric interaction couples the states as follows: $|0\rangle_a|0\rangle_b \leftrightarrow |1\rangle_a|1\rangle_b\leftrightarrow  |2\rangle_a|2\rangle_b \leftrightarrow\cdots$, $|0\rangle_a|1\rangle_b \leftrightarrow |1\rangle_a|2\rangle_b \leftrightarrow\cdots$ and $|1\rangle_a|0\rangle_b \leftrightarrow |2\rangle_a|1\rangle_b \leftrightarrow\cdots$.

As parametric interaction is turned on, some portion of the entanglement which is created in the initially occupied subspace $\{0110\}$ is transformed to other subspaces: $\{0220\}$ and $\{1221\}$. In consequence, one can suspect that it will be potentially possible to generate the following Bell-like states:
\begin{eqnarray}
|B\rangle_{0011}&=&\frac{1}{\sqrt{2}}\left(|0\rangle_a|0\rangle_b \pm |1\rangle_a|1\rangle_b\right),\nonumber\\
|B\rangle_{0110}&=&\frac{1}{\sqrt{2}}\left(|0\rangle_a|1\rangle_b \pm |1\rangle_a|0\rangle_b\right)\label{bell_0011_a}
\end{eqnarray}
and
\begin{eqnarray}
|B\rangle_{0022}&=&\frac{1}{\sqrt{2}}\left(|0\rangle_a|0\rangle_b \pm |2\rangle_a|2\rangle_b\right),\nonumber\\
|B\rangle_{1221}&=&\frac{1}{\sqrt{2}}\left(|1\rangle_a|2\rangle_b \pm |2\rangle_a|1\rangle_b\right).\label{bell_0011_b}
\end{eqnarray}
Obviously, if we assume that $s$ is small, we can expect that the entanglement in all subspaces considered here will be reduced. However, when $s$ becomes closed to the unity, the initial state becomes MES of the form $|B\rangle_{0011}=\frac{1}{\sqrt{2}}\left(|0\rangle_a|0\rangle_b \pm |1\rangle_a|1\rangle_b\right)$. Due to this fact and as a result of parametric excitation, the number of Bell-like states which could be generated decreases and only $|B\rangle_1=\frac{1}{\sqrt{2}}\left(|i\rangle_a|i\rangle_b \pm |j\rangle_a|j\rangle_b\right)$ ($i,j=0,1,2$ and $i\neq j$) could appear in the system. For such situation, the entanglement measured in subspaces ($\{0110\}$, $\{0220\}$ and $\{1221\}$) should  increase.

When the initial Werner-like  state involves Bell state   $|B\rangle_{0110}=\frac{1}{\sqrt{2}}\left(|0\rangle_a|1\rangle_b \pm |1\rangle_a|0\rangle_b\right)$ and  the parameter $s$ is sufficiently small, the situation is similar to that discussed above.  However, when $s$ increases, the states $|0\rangle_a|1\rangle_b$, $|1\rangle_a|0\rangle_b$ start to play more essential role. In consequence, one can expect that the entanglement present in the subspace $\{0220\}$ will be reduced, as there is no physical mechanism that can populate states $|0\rangle_a|2\rangle_b$, $|2\rangle_a|0\rangle_b$ and $|0\rangle_a|0\rangle_b$. Simultaneously, the entanglement will be created in the subspaces $\{0110\}$ and $\{1221\}$. In consequence, for this case the following Bell-like states could be potentially generated:
\begin{widetext}
\begin{eqnarray}
|B\rangle_{0110}&=&\frac{1}{\sqrt{2}}\left(|0\rangle_a|1\rangle_b \pm |1\rangle_a|0\rangle_b\right)\hspace*{0.5cm} |B\rangle_{1221}=\frac{1}{\sqrt{2}}\left(|1\rangle_a|2\rangle_b \pm |2\rangle_a|1\rangle_b\right)\\
|B\rangle_{0112}&=&\frac{1}{\sqrt{2}}\left(|0\rangle_a|1\rangle_b \pm |1\rangle_a|2\rangle_b\right) \hspace*{0.5cm} |B\rangle_{2110}=\frac{1}{\sqrt{2}}\left(|2\rangle_a|1\rangle_b \pm |1\rangle_a|0\rangle_b\right)
\label{bell_0110}
\end{eqnarray}
\end{widetext}

\section{Amplitude damping}
Amplitude damping processes are related to the energy losses. To describe damped system's evolution we shall use master equation, within Born and Markov approximations. That equation can be written as:
\begin{widetext}
\begin{eqnarray}   
\frac{d}{dt}\hat{\rho} &=&
 -i(\hat{H}\hat{\rho}-\hat{\rho}\hat{H})+ \gamma_a\left[
 \hat{a}\hat{\rho} \hat{a}^\dagger -\frac{1}{2}\left(\hat{\rho} \hat{a}^\dagger
 \hat{a} + \hat{a}^\dagger \hat{a}\hat{\rho} \right)\right]+
 \bar{n}_a\gamma_a\left[\hat{a}^\dagger\hat{\rho}
 \hat{a} + \hat{a}\hat{\rho} \hat{a}^\dagger - \hat{a}^\dagger \hat{a}\hat{\rho} - \hat{\rho} \hat{a}
 \hat{a}^\dagger \right]\nonumber\\
 &&+ \gamma_b\left[ \hat{b}\hat{\rho} \hat{b}^\dagger-\frac{1}{2}\left(\hat{\rho} \hat{b}^\dagger \hat{b}
  + \hat{b}^\dagger \hat{b}\hat{\rho}\right)\right] +\bar{n}_b\gamma_b\left[ \hat{b}^\dagger\hat{\rho} \hat{b}
  + \hat{b}\hat{\rho} \hat{b}^\dagger - \hat{b}^\dagger \hat{b}\hat{\rho} -\hat{\rho}\hat{b}
  \hat{b}^\dagger\right],\label{ampl_damp}
\label{master_1}
\end{eqnarray}
\end{widetext}
where $\gamma_{a,b}$ are damping constants and $\bar{n}_{a,b}$ denote mean number of photons and correspond to the non-zero temperature baths. We can solve this equation numerically and then, calculate negativities for various subspaces populated during the system's dynamics. As we can express all operators and density matrix in $n$-photon states basis, master equation can be writen in a form of a set of linear differential equations. Such equations can be solved numerically with use of standard procedures (for description of such procedures for instance, \cite{NR2007}). Negativity \cite{P96,HHH96} is entanglement monotone and for bipartite system is defined with use of the density matrix $\rho$ as in \cite{VW02}:
\begin{equation}
N(\rho )\,=\,\frac{\Vert \rho^{T_A}\Vert-1}{2},
\end{equation}
where $\rho^{T_A}$ is partial transpose of the density matrix and $\Vert X \Vert = Tr\sqrt{X^\dagger X}$ denotes trace norm. Since we shall discuss \textit{qubit-qubit} subsystems, such defined negativity will be equal to the unity for maximally entangled states and equal to zero for completely disentangled system.

\begin{figure}[htbp]
\centerline{\includegraphics[width=0.8\columnwidth]{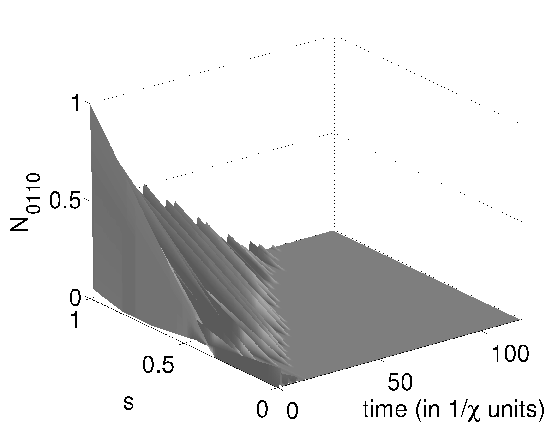}\hspace*{-1cm}\raisebox{5.5cm}{(a)}}
\centerline{\includegraphics[width=0.8\columnwidth]{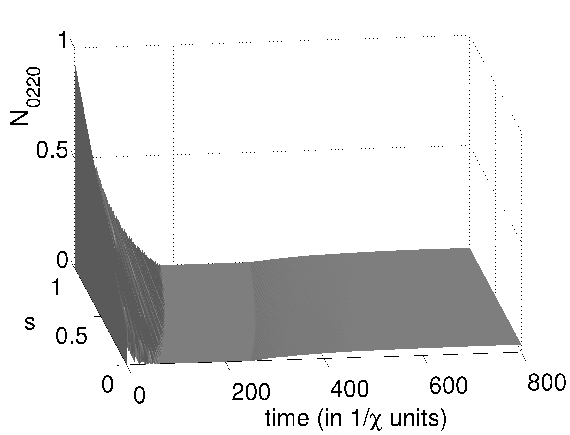}\hspace*{-1cm}\raisebox{5.5cm}{(b)}}
\caption{Time-evolution of the negativities $N_{0110}$ and $N_{0220}$ defined for subspaces $\{0110\}$ (subplot a) and $\{0220\}$ (subplot b), respectively, and for various values of the parameter $s$. Time is expressed in  units of $1/\chi$. Initial states are Werner-like states defined with use of $|B\rangle_1$ for $i=1$ as in (\ref{bell-like}). We assume amplitude damping reservoir and the parameters $g=0.6\chi$, $\gamma=0.005\chi$.}
\end{figure}

If we assume that the Bell state applied in the definition of the initial Werner state is of the form $|B\rangle_{0011}$ 
we can observe various features characteristic for the entanglement's evolution in damped systems.
In this paper we shall mostly concentrate on the effects of vanishing and reappearing of entanglement in the system. The disentanglement of initially entangled system, due to the interaction with the reservoir, usually occurs in asymptotic way.
If entanglement decays to zero in finite time one can say of entanglement sudden death \cite{ZHH01,YE04,ILZ07}.
This phenomenon can be observed in both Markovian and non-Markovian reservoirs as a consequence of the loss of the coherence between the qubits which is transferred to the modes of the cavity. In non-Markovian reservoirs such phenomenon can be even enhanced because of the memory effects \cite{MOP07}. When the entanglement reappears we deal with \textit{sudden entanglement birth}. Sudden death and/or its birth were widely reported for example, in systems of atoms forming entangled qubits \cite{FT06,FT08} or  harmonic oscillators \cite{BF06}. 

It should be emphasized that sudden vanishing of the entanglement, and its rebirth, can appear in systems in which damping processes are neglected. For such  cases the entanglement ''flows`` from one subspace of the states (or from one subsystem) to another one, leading to sudden disappearing of the entanglement in given subspace (or subsystem). Such phenomena are sometimes interpreted as sudden entanglement death or birth, but they differ in their nature from those induced by interactions with external bath. Sometimes the both classes of phenomena can be observed in one system. For instance, such situation was discussed in \cite{KL09}, where the model involving two Kerr-like oscillators was considered.

 We shall concentrate on the two two-qubit subspaces: $\{0110\}$ and $\{0220\}$.  For the entanglement within subspace  $\{0110\}$  one can observe its decay in a finite time for all values of $s$ (Fig.2a). For the subspace $\{0220\}$ (Fig.2b) we observe different behavior. We see that for all initial states (all values of the parameter $s$) after some entanglement disappearance in finite time, there is a large time period of no entanglement at all. Then it comes back again to this subspace. As it was mentioned earlier, these effects are sometimes referred to as \textit{entanglement death} and its \textit{rebirth}.

\begin{figure}[htbp]
\centerline{\includegraphics[width=0.8\columnwidth]{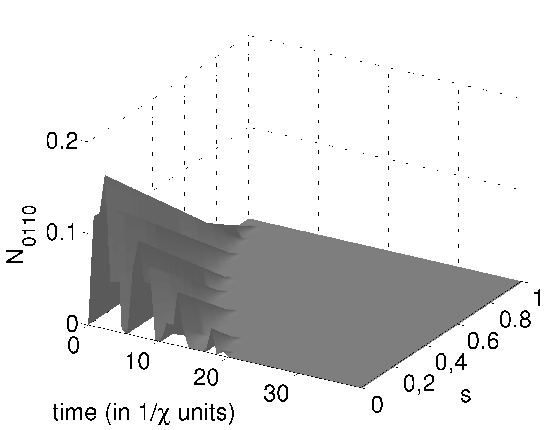}\hspace*{-1cm}\raisebox{5.7cm}{(a)}}
\centerline{\includegraphics[width=0.8\columnwidth]{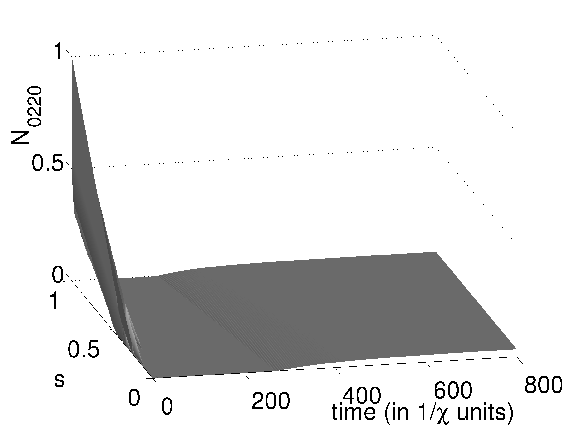}\hspace*{-1cm}\raisebox{5.5cm}{(b)}}
\centerline{\includegraphics[width=0.8\columnwidth]{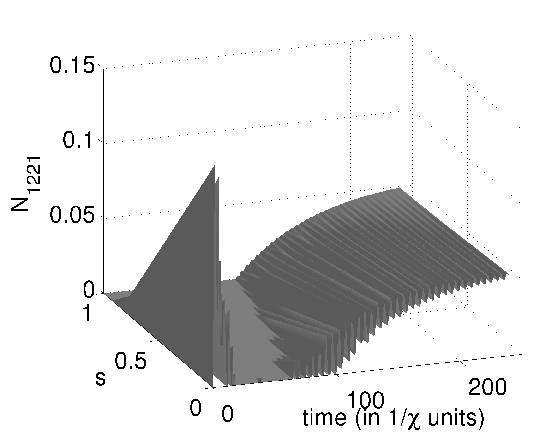}\hspace*{-1cm}\raisebox{5.7cm}{(c)}}
\caption{The same as in Fig.2 but for initial Werner-like state defined by $|B\rangle_2$ for $i=2$ as in (\ref{bell-like}). The negativities are: $N_{0110}$ -- (a), $N_{0220}$ -- (b) and $N_{1221}$ -- (c).}
\end{figure}
For initial Werner-like states defined with use of the state: $|B\rangle_{0220}$ 
the character of the two-qubit entanglement decay differs considerably in its character from that presented in Fig.2.  Such behavior is a result of the fact that initially two-photon states are populated. Then, due to the parametric interaction and amplitude decay other states (one-photon and vacuum states) are involved in the system's evolution, as well. The probabilities corresponding to all those states evolve in different way from those discussed in previous case. Thus, in Fig.3a one can see that  for the  subspace $\{0110\}$  the more the mixed state is presented in the initial Werner-like state, the more  entanglement is created. We observe this effect even though the oscillations of the negativity $N_{0110}$ are of small amplitude and disappear rather fast. 

For subspace $\{0220\}$ we can obtain more entanglement (larger negativity) than for the  subspace $\{0110\}$ (Fig.3b). For $0\leq s \leq 0.15$, when the initial  state is separable, we can see that entanglement within this subspace is ''born'' after a long period of time. So, the birth of entanglement can be in this subspace delayed. But for all other initial states (which are initially entangled) the behavior of negativity $N_{0220}$  has the same character as for the situation presented in Fig.2b  -- entanglement death after some time of decaying oscillations and its revival after an additional long period of time, can be observed. Similar features are present in subspace $\{1221\}$, as well (Fig.3c), but for the cases when $0\leq s \leq 0.93$. Moreover, the values of negativity $N_{1221}$ do not exceed those of $N_{0110}$.
For the initial states which are close to the Bell state $|B\rangle_{0220}$
there is no possibility to obtain non-zero values for $N_{1221}$ as a result of parametric interaction.
This interaction itself cannot transfer the population from the states forming $|B\rangle_{0220}$ to those defining the subspace $\{1221\}$. It can only be possible for the system interacting with amplitude damping reservoir, when initially populated states decay at first to the states from other subspaces, and only after that the parametric interaction is able to populate states from $\{1221\}$. That is why we can observe the delayed birth of entanglement shown in Fig.3b.

The features which we pointed out in the previous considerations (entanglement death, its birth or revival) can be explained with use of the relations between the populations and coherences of the adequate two-qubit density matrices. Such matrices have in general the following form:
\begin{equation}   
 \hat{\rho}^{\rm red} = \left(
   \begin{array}{cccc}
   a_{11} & a_{12} & a_{13} & a_{14} \\
   a_{21} & a_{22} & a_{23} & a_{24} \\
   a_{31} & a_{32} & a_{33} & a_{34} \\
   a_{41} & a_{42} & a_{43} & a_{44}
  \end{array} \right) ,
  \end{equation}
where $a_{kl}$ are either the populations (for $k=l$) or coherences when $k\neq l$.
It appears that for the system under considerations in a specific situations (defined by the form of the initial state) the number of nonzero matrix elements is significantly reduced. In fact, density matrices describing our model are of the following forms:
\begin{eqnarray}   
 \hat{\rho}^{\rm red}_{A_1}& =& \left(
   \begin{array}{cccc}
   a_{11} & 0 & 0 & a_{14} \\
   0 & a_{22} & 0 & 0 \\
   0 & 0 & a_{33} & 0 \\
   a_{41} & 0 & 0 & a_{44}
  \end{array} \right)\nonumber\\
  \hat{\rho}^{\rm red}_{A_2} &=& \left(
   \begin{array}{cccc}
   a_{11} & 0 & 0 & a_{14} \\
   0 & a_{22} & a_{23} & 0 \\
   0 & a_{32} & a_{33} & 0 \\
   a_{41} & 0 & 0 & a_{44}
  \end{array} \right) \nonumber\\
\label{matr_ampl}
\end{eqnarray}  
where $\hat{\rho}^{\rm red}_{A_1}$ corresponds to the Bell-like state $|B\rangle_1=\frac{1}{\sqrt{2}}\left(|0\rangle_a|0\rangle_b \pm |i\rangle_a|i\rangle_b\right)$ applied to the definition (\ref{wer1}), whereas $\hat{\rho}^{\rm red}_{A_2}$ corresponds to $|B\rangle_2=\frac{1}{\sqrt{2}}\left(|0\rangle_a|i\rangle_b \pm |i\rangle_a|0\rangle_b\right)$.

The matrices (\ref{matr_ampl}) are of $X$-type and hence, the conditions for obtaining zero-valued negativities can be easily derived. After performing partial transposition of the density  matrices we can find the expressions for their eigenvalues and then, the conditions for their negative values. Thus, we obtain:
\begin{eqnarray}  
 \lambda_{1_{A_1}}&=&a_{11}, \nonumber\\
 \lambda_{2_{A_1}}&=&\frac{1}{2}\left[a_{22}+a_{33}-\sqrt{\left(a_{22}-a_{33}\right)^2+4a_{14}a_{41}}\right],
  \nonumber\\
 \lambda_{3_{A_1}}&=&\frac{1}{2}\left[a_{22}+a_{33}+\sqrt{\left(a_{22}-a_{33}\right)^2+4a_{14}a_{41}}\right],
 \nonumber\\
 \lambda_{4_{A_1}}&=&a_{44}\label{eig1}.
\end{eqnarray}
for density matrix  $\hat{\rho}^{\rm red}_{A_1}$. The equality $a_{22}a_{33}=a_{14}a_{41}$ gives the border condition for occuring entanglement death and its revival within two-qubit subspace described by this matrix. 

From the other side, the eigenvalues:
\begin{eqnarray}  
 \lambda_{1_{A_2}}&=&\frac{1}{2}\left[a_{11}+a_{44}-\sqrt{a_{11}^2+a_{44}^2-2(a_{11}a_{44}-2a_{23}a_{32})}\right], \nonumber\\
 \lambda_{2_{A_2}}&=& \frac{1}{2}\left[a_{11}+a_{44}+\sqrt{a_{11}^2+a_{44}^2-2(a_{11}a_{44}-2a_{23}a_{32})}\right],\nonumber\\
 \lambda_{3_{A_2}}&=&\frac{1}{2}\left[a_{22}+a_{33}-\sqrt{a_{22}^2+a_{33}^2-2(a_{22}a_{33}-2a_{14}a_{41})}\right], \nonumber\\
 \lambda_{4_{A_2}}&=&\frac{1}{2}\left[a_{22}+a_{33}+\sqrt{a_{22}^2+a_{33}^2-2(a_{22}a_{33}-2a_{14}a_{41})}\right].\nonumber\\
\label{eig2}
\end{eqnarray}
of $\hat{\rho}^{\rm red}_{A_2}$ allow for deriving the equations $a_{22}a_{33}=a_{14}a_{41}$ and  $a_{11}a_{44}=a_{23}a_{32}$,  as possible conditions for occurrence of entanglement death and its revival within two-qubit subspace described by $\hat{\rho}^{\rm red}_{A_2}$.
In fact, the eigenvalue $\lambda_{3_{A_2}}$ has only positive values and the only condition for obtaining $N=0$ is that which follows from the negative value of $\lambda_{1_{A_2}}$.

As an example we present: the coherences, adequate populations and negativity which corresponds the entanglement within the subspace described by the apropriate two-qubit matrix -- see Fig.4.
\begin{figure}[htbp]
\centerline{\includegraphics[width=1.1\columnwidth]{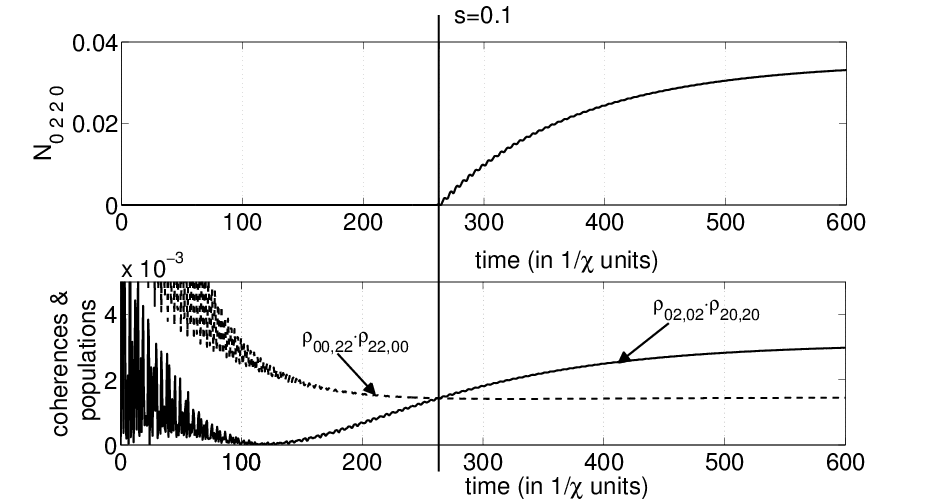}}
\caption{Time-evolution of the negativity $N_{0110}$ (top), populations and coherences (bottom). Initial Werner-like state is defined by $|B\rangle_2$ for $i=2$ as in eqn. (\ref{bell-like}). Time is scaled in $1/\chi$ units and me for  Werner-like state defined by .The parameter $s=0.1$ $g=0.6\chi$, $\gamma=0.005\chi$, amplitude damping reservoir. }
\end{figure}

It can be clearly seen that in fact we can correlate the occurence of nonzero values of negativity with relation $\rho_{02,02}\rho_{20,20}>\rho_{00,22}\rho_{22,00}$. This correlation originates from the fact that in the process discussed here the dynamics of that type which can be described by the X, or X-type matrices corresponding to the two-qubit subspaces only.
In the system of two anharmonic oscillators coupled by parametric process and decaying in the Markovian reservoir the whole system is at least 2-qutrit one. The entanglement however is traced within the 2-qubit subspaces of the whole system. It is known, that coherences between the qubits are responsible for observing entanglement \cite{W98,FT08}. At the begining of the interaction, the states $|1\rangle_a|1\rangle_b$ and $|2\rangle_a|2\rangle_b$ are quickly populated, but due to the damping process populations of these states decay and in the long time limit population of the state  $|0\rangle_a|0\rangle_b$ increases.
The main reason for observing vanishing and reapearing of the entanglement in discussed subspaces is fact that the interaction with external bath reduces populations for higher $n$-photon states and simultaneously, decreses coherences for all of them. At the same time, external excitation "pumps" the coherences for the states belonging to all subspaces. These two mechanisms lead to the changes of the ratios between the coherences and populations observed in various subspaces. In consequence, populations for lower states more likely exceed coherences in subspaces corresponding to the vacuum and one-photon states, leading to the disappearing of the entanglement. Simultaneously, for sub-spaces involving two-photon states the coherences can dominate over the populations appearing there and hence, the entanglement can (re)appear.
Finally, the system reaches some kind of balance between the pumping and damping processes. Those effects we can see  in Fig.4, where the populations corresponding to the states defined  in $\{0110\}$ subspace increase and become higher than coherences. In consequence, entanglement death is observed in this subspace. For the other subspace $\{0220\}$ the populations initially increase, and after some time they reach values higher than values of coherences. It causes entanglement death but finally, damping process reduces populations and they are exceeded by coherences again (for the analysis of such effects see also \cite{KLP11}). 

\section{Phase damping}
For the second model of damping -- phase damping, there is no loss of energy and populations of the states represented by diagonal matrix elements do not decay. For such cases decoherence effects and hence, entanglement losses due to the decay of other than diagonal matrix elements can be observed. The process of the decoherence induced by the phase damping reservoir is related to random changes of relative phases of superposed states during system's time-evolution. Such behavior is also interpreted as a kind of scattering process.

For phase damping processes the master equation corresponding to the system -- external environment interaction can be written as:
\begin{eqnarray}
\frac{d}{dt}\hat{\rho} &=&
 -i(\hat{H}\hat{\rho}-\hat{\rho}\hat{H})\\
 &+&\frac{1}{2}\gamma_a\left(\bar{n}_a+1\right)\left[2\hat{a}^\dagger\hat{a}\hat{\rho}\hat{a}^\dagger\hat{a}-
 \left(\hat{a}^\dagger\hat{a}\right)^2\hat{\rho}-\hat{\rho}\left(\hat{a}^\dagger\hat{a}\right)^2\right]\nonumber \\
 &+&\frac{1}{2}\gamma_b\left(\bar{n}_b+1\right)\left[2\hat{b}^\dagger\hat{b}\hat{\rho}\hat{b}^\dagger\hat{b}-
 \left(\hat{b}^\dagger\hat{b}\right)^2\hat{\rho}-\hat{\rho}\left(\hat{b}^\dagger\hat{b}\right)^2\right]\nonumber\label{phase_meq}
\end{eqnarray}
For the density matrices determined by this equation we see that they are  $4\times 4$ (two-qubit) matrices. We can find that for this case density matrices are of the previously discussed X-type shown in (\ref{matr_ampl}). Moreover, they can also take one of the following forms:
\begin{equation}
\hat{\rho}^{\rm red}_{Ph_1} =\left(
   \begin{array}{cccc}
   a_{11} & 0 & 0 & a_{14} \\
   0 & 0 & 0 & 0 \\
   0 & 0 & 0 & 0 \\
   a_{41} & 0 & 0 & a_{44}
  \end{array} \right) ,
  \hspace*{0.3cm}
  \hat{\rho}^{\rm red}_{Ph_2} = \left(
   \begin{array}{cccc}
   0 & 0 & 0 & 0\\
   0 & a_{22} & a_{23} & 0 \\
   0 & a_{32} & a_{33} & 0 \\
   0 & 0 & 0 & 0
  \end{array} \right) 
 \label{matr_ph}
\end{equation}
Analogously to the case of amplitude damping, we can find eigenvalues for density matrix $\hat{\rho}^{\rm red}_{Ph_1}$. They are:
\begin{eqnarray} 
\lambda_{1_{ph_1}}&=&-\sqrt{a_{14}a_{41}}\nonumber\\
\lambda_{2_{ph_1}}&=&\sqrt{a_{14}a_{41}}\nonumber\\
\lambda_{3_{ph_1}}&=&a_{11}\nonumber\\
\lambda_{4_{ph_1}}&=&a_{44}
\end{eqnarray}
or 
\begin{eqnarray} 
\lambda_{1_{ph_2}}&=&-\sqrt{a_{23}a_{32}}\nonumber\\
\lambda_{2_{ph_2}}&=&\sqrt{a_{23}a_{32}}\nonumber\\
\lambda_{3_{ph_2}}&=&a_{22}\nonumber\\
\lambda_{4_{ph_2}}&=&a_{33}
\end{eqnarray}
for the matrix $\hat{\rho}^{\rm red}_{Ph_2}$.
It can be seen that negative eigenvalue (which is necessary for obtaining negativity larger than zero) can appear only when $\sqrt{a_{14}a_{41}}$ or $\sqrt{a_{23}a_{32}}$ are larger than zero. That means that in fact it is sufficient condition to get a two-qubit entanglement within discussed subspaces when density matrix elements (appearing in formulas determining $ \lambda_{1_{ph_1}}$ or $\lambda_{1_{ph_2}}$) have nonzero off-diagonal values.

Matrices of the type  $\hat{\rho}^{\rm red}_{Ph_1}$,  appearing in (\ref{matr_ph}), can be obtained only when the following Bell-like states are used for the definition of initial Werner-like state:
\begin{itemize}
\item $|B\rangle_{0110}=\frac{1}{\sqrt{2}}\left(|0\rangle_a|1\rangle_b \pm |1\rangle_a|0\rangle_b\right)$ and $s\neq 1$ in subspace $\{0220\}$,
\item $|B\rangle_{0220}=\frac{1}{\sqrt{2}}\left(|0\rangle_a|2\rangle_b \pm |2\rangle_a|0\rangle_b\right)$ and $s\neq 1$ in subspace $\{0110\}$,
\item $|B\rangle_{0011}=\frac{1}{\sqrt{2}}\left(|0\rangle_a|0\rangle_b \pm |1\rangle_a|1\rangle_b\right)$ and all values of $s$ in subspace $\{0220\}$, or for $s=1$ in subspace $\{0110\}$,
\item $|B\rangle_{0022}=\frac{1}{\sqrt{2}}\left(|0\rangle_a|0\rangle_b \pm |2\rangle_a|2\rangle_b\right)$ and all values of $s$ in subspace $\{0110\}$, or for $s=1$ in subspace $\{0220\}$.
\end{itemize}
On the other hand, matrices of the type  $\hat{\rho}^{\rm red}_{Ph_2}$  can be obtained for:
\begin{itemize}
\item $|B\rangle_{0110}=\frac{1}{\sqrt{2}}\left(|0\rangle_a|1\rangle_b \pm |1\rangle_a|0\rangle_b\right)$ for subspace $\{0110\}$,
\item $|B\rangle_{0220}=\frac{1}{\sqrt{2}}\left(|0\rangle_a|2\rangle_b \pm |2\rangle_a|0\rangle_b\right)$ for subspace $\{0220\}$.
\end{itemize}
For other situations the matrices  have one of the forms discussed in previous section and shown in (\ref{matr_ampl}).

We can compare the conditions for obtaining zero negativity with those for entanglement death in phase damping reservoir given by Huang and  S. Zhu in \cite{HZ07a}. According to their statement it is only possible to observe entanglement death in such phase damping reservoir in two-qubit system in which the initial density matrix describing the system has all nonzero diagonal elements \textit{i.e.}
\begin{equation}
a_{11}a_{22}a_{33}a_{44}\neq 0
\label{death_cond}
\end{equation}
This condition is not fulfilled for all the initial states for which our system is described by one of the matrices given by (\ref{matr_ph}). Fig.5 shows the time-evolution of the negativity for such initial states.
In Fig.5a we see that for all values of $s\neq 1$ entanglement in subspace $\{0110\}$ oscillates and then, disappears in finite time ---  for all values of $s$ density matrix for two-qubit subsystem is of the form defined by $\hat{\rho}^{\rm red}_{A_2}$  and for all  states considered there the condition (\ref{death_cond}) is fulfilled. Only for $s=1$ the system is described by the matrix $\hat{\rho}^{\rm red}_{Ph_2}$ and for that case the oscillations have an asymptotic decay.  Fig.5b shows analogous situation but for the negativity $N_{0220}$. Again, when $s\neq 1$ the system is described by the matrix which satisfies the condition (\ref{death_cond})and the entanglement within this subspace disappears in finite time. For the case when $s=1$ the state of the system is such that it can be defined by the matrix  $\hat{\rho}^{\rm red}_{Ph_2}$ and in consequence,  the asymptotic decay will be present.
\begin{figure}[htbp]
\centerline{\includegraphics[width=0.8\columnwidth]{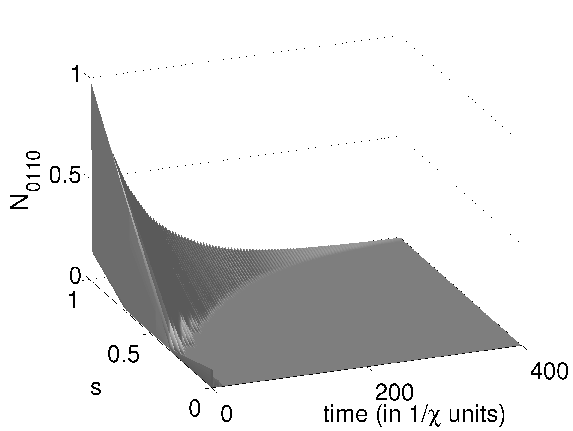}\hspace*{-1cm}\raisebox{5cm}{(a)}}
\centerline{\includegraphics[width=0.8\columnwidth]{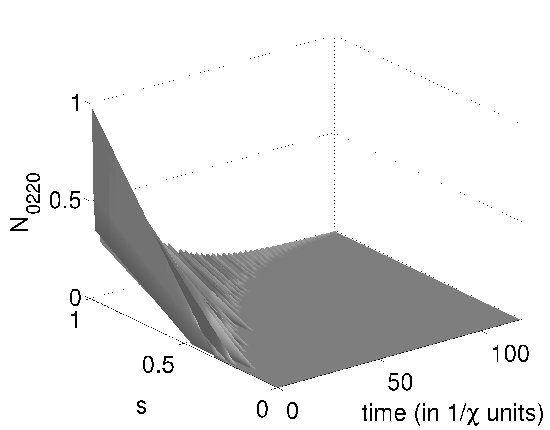}\hspace*{-1cm}\raisebox{5cm}{(b)}}
\caption{Time-evolution of two-qubit negativities $N_{0110}$ (a) and $N_{0220}$ (b) for phase damped system and for all possible values of $s$. Initial states are: (a) -- Werner-like state defined by $|B\rangle_2$ for $i=1$ in (\ref{bell-like}) and (b) --  Werner-like state defined by $|B\rangle_2$ for $i=2$ in (\ref{bell-like}). Additionally, $g=0.6\chi$ and $\gamma=0.005\chi$. Time is scaled in $1/\chi$ units.}
\end{figure}

It is also worth mentioning that in \cite{K12} the discussion for the present model in phase damping reservoir was performed  for maximally entangled initial states only, and that corresponds to one of the matrices (\ref{matr_ph}). For such matrices the condition (\ref{death_cond}) was not fulfilled, so that it was not possible to observe entanglement death in situations considered there  --  only oscillations with asymptotically decaying amplitude were present.

\section{Conclusions}
We discussed the model of parametrically pumped Kerr-like coupler concentrating on the the influence of the initial Werner-like state on the system's dynamics. In particular, we found that the model can behave as NQS, similarly as for that previously discussed in \cite{KLP11},  where Bell-like states generation from initial Fock states was considered.  In this paper we showed that when the same system starts its evolution from Werner-like states, it can exhibit various, not discussed in \cite{KLP11} features. We discussed the possibility of creation of entanglement for cases of all possible mixtures of MEPS and maximally mixed state, showing that even though the initial state is a mixed one (or mixed with some small addition of the entangled state), it is possible to truncate the number of states that are relevant in the system's dynamics in such a way that  two-qubit pure entangled state can be generated. Moreover, for some values of the parameters describing Werner-like initial states, delayed entanglement creation can be observed.

We also analyzed the process of disentanglement  which can be achieved in two types of  damping processes: amplitude  and phase ones. For amplitude damping, we have confirmed appearance of the features which have been reported in \cite{KLP11}: entanglement death in one 2-qubit subspace ($\{0110\}$), entanglement death and revival in the subspace ($\{0220\}$) and entanglement stay at a certain value in the subspace ($\{1221\}$). Thus, we have shown that these features  are not characteristic  for initial pure $n$-foton states only. We  have  confirmed this fact for a large group of initial Werner-like states,  defined by the Bell state $|B\rangle_1$. What is interesting, if we assume that the state $|B\rangle_2$ defines initial Werner-like state, it is possible to observe a \textsl{delayed entanglement birth}  -- after some initial period of time when the system stays in nonentangled state, some portion of the entanglement is produced. It can be observed when initial Werner-like state is not entangled and $s\in\langle 0;0.15)$. 

Moreover, we showed that when  phase damping processes act on the system, two ways of disentangling can be observed. Depending on the value of the parameter describing initial Werner-like state, the entanglement can disappear with oscillating and asymptotically decaying amplitude, or alternatively, it vanishes suddenly  after some finite period of time. For the latter case, we have identified the conditions for observing  entanglement death. We pointed out that the reason for such behavior, which has its origin in the relation of the coherences to populations of the adequate two-qubit density matrices, is consistent with the condition (given earlier in \cite{HZ07a}).

\section*{Acknowledgments}

A. K-K would like to thank NCN grant No DEC-2011/03/B/ST2/01903  for the support. 
The authors wish to thank prof. Jan Pe\v{r}ina Jr. for his valuable discussions and suggestions.


\end{document}